\documentclass[]{article}
\usepackage{cite}
\usepackage{amsmath,amssymb,amsfonts}
\usepackage{algorithmic}
\usepackage{graphicx}
\usepackage{subcaption}
\usepackage{textcomp}
\usepackage{xcolor}
\usepackage{multirow}
\usepackage{adjustbox}
\usepackage{booktabs}
\usepackage{tabularx}
\usepackage{hhline}
\usepackage{url}
\usepackage[square,numbers,sort&compress]{natbib}

\usepackage{fancyhdr}
\let\origtitle\title 
\renewcommand{\title}[1]{\lfoot{\textit{#1}}\origtitle{\textbf{#1}}}
\renewcommand{\sectionmark}[1]{\markboth {}{}}

\date{}

\title{TrackThinkDashboard: Understanding Student Self-Regulated Learning in Programming Study}

\begin{document}
\maketitle
\thispagestyle{fancy}
\centering

\author{
Ko Watanabe \footnote{ko.watanabe@dfki.de}, 
Yuki Matsuda \footnote{yukimat@okayama-u.ac.jp}, 
Yugo Nakamura \footnote{y-nakamura@ait.kyushu-u.ac.jp}, \\
Yutaka Arakawa \footnote{arakawa@ait.kyushu-u.ac.jp}, and 
Shoya Ishimaru \footnote{ishimaru@omu.ac.jp}} \\
\thanks{
$^1$ German Research Center of Artificial Intelligence (DFKI GmbH), \\
$^2$ Okayama University, 
$^3$$^4$ Kyushu University, 
$^5$ Osaka Metropolitan University 
}

\abstract{
In programming education, fostering self-regulated learning (SRL) skills is essential for both students and teachers. This paper introduces \textit{TrackThinkDashboard}, an application designed to visualize the learning workflow by combining web browsing and programming logs in one unified view. The system aims to (1) help students monitor and reflect on their problem-solving processes, identify knowledge gaps, and cultivate effective SRL strategies, and (2) enable teachers to identify at-risk learners more effectively and provide targeted, data-driven guidance. We conducted a study with 33 participants (32 male, one female) from Japanese universities—some with prior programming instruction and some without—to explore differences in web browsing and coding patterns. The dashboards revealed multiple learning approaches (e.g., try-and-error, try-and-search, and more) and highlighted how domain knowledge influenced overall activity flow. We discuss how this visualization can be used continuously or in one-off experiments, the privacy considerations involved, and opportunities for expanding data sources for richer behavioral insights.
}

\section{Introduction}
\label{section:Introduction}
Programming education has emerged as a vital component of modern professional skill development. To meet the demands of an increasingly technology-driven world, educators must provide hands-on learning experiences that nurture students' self-regulated learning (SRL) skills~\cite{zimmerman2001self}. SRL empowers learners to take charge of their educational journey through goal setting, self-monitoring, reflection, and refinement, ultimately enhancing both academic performance and long-term knowledge retention.

\begin{figure*}[t] 
  \centering 
  \includegraphics[width=0.9\columnwidth]{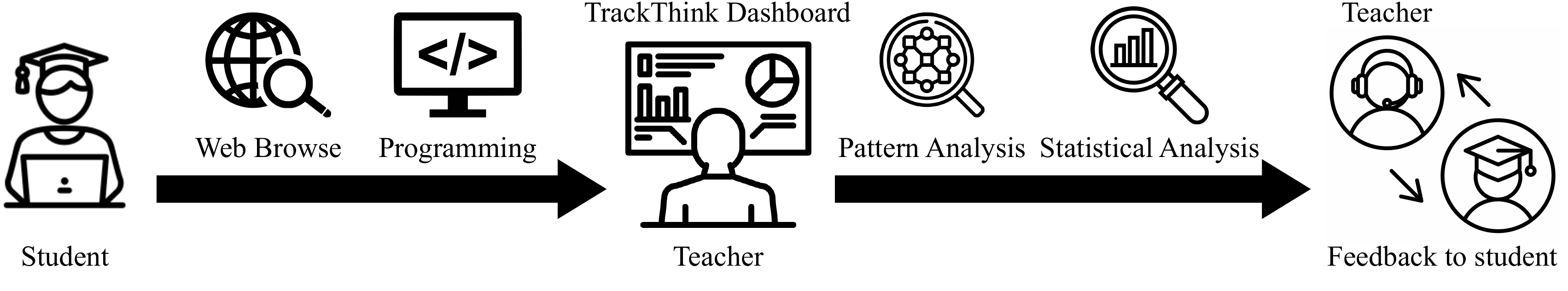} 
  \caption{Overall workflow of \textit{TrackThinkDashboard}.} 
  \label{fig:workflow_process_trackthink_dashboard} 
\end{figure*}

To enhance SRL in programming education, we present the \textit{TrackThinkDashboard}, an innovative application designed to visualize SRL activity behaviors. The application aims to empower students and teachers by integrating web browsing and programming activities into a unified view. \citeauthor{prather2020what} mentioned that many novices do not have well-developed content knowledge or cognitive control in programming, so they lack even a basic understanding of their progress through the programming problem-solving process. Visualizing whole programming and web browsing activities can help students gain deeper insights into their learning workflows and decision-making processes, fostering a more self-aware and practical approach to learning~\cite{prather2020what}.

While previous systems like SearchBar~\cite{Morris2008}, popHistory~\cite{Carrasco2017}, and \textit{TrackThinkTS}~\cite{makhlouf2022} have focused on collecting web browsing logs, and tools such as Log++~\cite{marron2018} and Projection Boxes~\cite{Lerner2020} have captured programming activity logs, these studies have not addressed the integration and classification of behaviors across both domains. We chose to combine web browsing and programming logs because, in many programming tasks, acquiring knowledge (through web searches, documentation, or Q\&A sites) is tightly coupled with the application of knowledge (writing, compiling, and revising code)~\cite{prather2020what}. Observations in programming education~\cite{brandt2009two, prather2020what, watanabe2022how} confirm that novice and intermediate learners rely heavily on real-time web searches, especially for syntax help or debugging. By syncing these logs in a single visualization, we give both students and teachers a complete picture of the problem-solving loop—making it clear when a student repeatedly returns to the same resource or whether they compile the code multiple times before searching.

The application serves both students and teachers by tracking and visualizing transitions between web searches and coding tasks. For students, these visualizations facilitate meta-cognitive awareness~\cite{loksa2022metacognition}, enabling them to pinpoint knowledge gaps, optimize resource usage, and develop more structured SRL strategies. In parallel, the dashboard equips teachers with detailed insights into student behavior, such as prolonged resource searching, recurring coding errors, or inefficient task-switching—allowing for targeted, data-driven interventions. The platform fosters collaboration and strengthens the overall learning experience for both parties by prompting discussions around specific learning behaviors and strategies.

This paper proposes an application that leverages web browsing and programming activities to visualize students' SRL workflow in programming learning. As shown in Figure~\ref{fig:workflow_process_trackthink_dashboard}, we use a web browsing activity logger~\cite{makhlouf2022} and a programming IDE to collect logs. The \textit{TrackThinkDashboard} synchronously visualizes two activities, offering an intuitive GUI (Graphical User Interface) that allows users to visualize their activities as a structured flowchart. Our contributions are as follows: 

\begin{enumerate} 
  \item \textbf{A unified visualization of web browsing and programming}: We introduce an application that combines web browsing and programming activities as an all-in-one flowchart.
  \item \textbf{We discover the pattern of SRL behavior through web browsing and programming activity}: Using our application, we identify how students transition between web browsing and programming tasks, revealing an understanding of the patterns of SRL workflow.
\end{enumerate}

\section{Related Work}
\subsection{SRL in Programming}
SRL is the cyclical process of planning, monitoring, and evaluating one's cognitive, meta-cognitive, and behavioral strategies to achieve learning goals~\cite{zimmerman2001self}. In programming, SRL encompasses how learners anticipate and design solutions, continuously monitor and debug their code, and then reflect on errors and outcomes to refine their mental models. Research by \citet{prather2020what} indicates that many novices struggle to gauge their progress, suggesting that visualizing one's workflow—such as errors, code searches, and version history—can reinforce meta-cognitive skills and help students develop more accurate mental models of programming.

More recently, several studies have shown how SRL and meta-cognition theories can be applied specifically to programming courses. For instance, \citet{loksa2022metacognition} present a systematic overview of SRL frameworks, illustrating how they inform both research and practice. Meanwhile, \citet{silva2024what} identify a broad spectrum of regulatory strategies that learners employ (e.g., time management, motivation, and planning), offering new insights into how SRL unfolds in the process of writing code. Collectively, these findings highlight the most common pitfalls novices face—such as challenges in debugging and sustaining motivation—and propose domain-specific scaffolding to strengthen students’ self-monitoring and self-evaluation skills in programming.

In summary, building on these studies, we argue that cultivating SRL in programming should focus on enhancing meta-cognitive awareness of each learner’s workflow, including how they write and troubleshoot code, as well as how they search for information. By visualizing the entire process—from planning and coding to reflection and debugging—novices can more effectively self-monitor and self-evaluate, ultimately becoming more self-regulated and proficient programmers.

\subsection{Web Browsing Activity Log Collection}
Several tools have been developed to log and analyze web browsing activities. Mermite~\cite{wong2007} lets users store multiple web resources and create mashups without needing programming skills. Users can organize content, like flight prices or publication years, in any order. SearchBar~\cite{Morris2008} saves browsing and query histories along with user ratings and notes, making it easier to highlight important actions, such as favorite websites or frequently used queries. This approach helps users visualize their browsing behavior and even share patterns with others. PopHistory~\cite{Carrasco2017} presents web history as a bubble chart, where bubble sizes indicate frequently visited sites. LogCanvas~\cite{xu2018} collects web search histories and visualizes them as knowledge graphs, helping users see connections between pages rather than simply listing URLs. Lastly, \textit{TrackThinkTS}~\cite{makhlouf2022, watanabe2023trackthink} is a browser extension inspired by TrackThink~\cite{nagano2017}. It logs tab and window operations with timestamps and exports data in CSV format, simplifying log analysis. These tools underscore the value of web browsing logs for both personal and shared insights.

\subsection{Programming Activity Log Collection}
Several tools have been developed to log and analyze programming activities. Log++~\cite{marron2018} captures logging results in JSON format, focusing on optimizing logging statements rather than recording compilation histories or results. Projection Boxes~\cite{Lerner2020} provides a summary of compilation results and timestamps, displaying code statements in a box format, but it’s not designed to collect detailed compile logs. \textit{C2Room} is an online programming IDE that logs timestamps, compiled code, and results. It also allows users to set programming tasks, recording task IDs and timestamps, making it easy to track code progress and outcomes. Similarly, WEVL~\cite{su14138084} supports online programming environments with comparable functionality. Lastly, Log-it~\cite{Jiang2023} is a Visual Studio Code extension that visualizes programming workflows, enabling users to easily reference historical code patterns through its visual interface.

\subsection{Visualize Web Browsing and Programming}
The integration of web browsing and programming logs has drawn attention in recent research to better understand data interactions. Prompter~\cite{ponzanelli2016} is an IDE plugin that retrieves relevant Stack Overflow discussions, ranks their relevance using a multifaceted model, and displays them to developers within the Eclipse IDE. 
A study involving 33 developers reported a 74\% positive response rate~\cite{ponzanelli2016}.
This concept was later expanded in Libra~\cite{ponzanelli2017}, which provides comprehensive web resource recommendations directly within the IDE. CrossRec~\cite{phuong2020} recommends third-party libraries by analyzing attached code. \citeauthor{brandt2009two} explored how developers use online resources for technical problem-solving~\cite{brandt2009two}, revealing that programmers engage in \textit{just-in-time} learning and use online materials for both acquiring new skills and clarifying existing knowledge. Additionally, \citeauthor{watanabe2022how} investigated the estimation of programming domain knowledge from web and programming log data~\cite{watanabe2022how}. By categorizing participants as domain experts or novices, the study achieved a prediction accuracy of 0.95 using Random Forest, highlighting significant behavioral differences between novice and expert programmers in web browsing and coding.

In summary, while existing works provide valuable insights into integrating web browsing and programming logs, none offer a clear visualization of the learning path in programming.

\section{Methodology}
\label{section:methodology}
We use two applications for logging web browsing and programming activity to visualize both activities synchronously. 
This section explains data source, fusion, and visualization approaches for understanding SRL.

\begin{table}[t!]
  \centering
  \caption{Web browsing log collected by \textit{TrackThinkTS}.}
  \renewcommand{\arraystretch}{1.0}
  \small
  \begin{tabular}{l p{5.0cm}}\hline
    Column Name                  & Description \\ \hline
    UserID                       & Participant's user ID. \\
    UserAction                   & User action (e.g., tab, scroll, copy). \\
    date                         & Log timestamp in UNIXTIME. \\
    Tab\_URL                     & URL of the accessed page. \\
    Tab\_Title                   & Title of the accessed page. \\
    Tab\_BodyText                & Body text of the accessed page. \\
    ClipboardCopy                & Copied text content. \\
    Scroll\_YAxisSpeed           & Vertical scroll speed. \\
    Scroll\_VisibleText          & Visible text after scroll. \\
    Scroll\_ViewPort\_XScroll    & Horizontal scroll viewport. \\
    Scroll\_ViewPort\_YScroll    & Vertical scroll viewport. \\
    Scroll\_XScrollRate          & Horizontal scroll percentage. \\
    Scroll\_YScrollRate          & Vertical scroll percentage. \\
    ViewPortWidth                & Viewport width. \\
    ViewPortHeight               & Viewport height. \\
    DocWidth                     & Document width. \\
    DocHeight                    & Document height. \\ \hline
  \end{tabular}
\label{tab:table_TrackThinkTS}
\end{table}

\begin{table}[t!]
  \centering
  \caption{Programming log collected by \textit{C2Room}.}
  \renewcommand{\arraystretch}{1.0}
  \small
  \begin{tabular}{l l}\hline
    Column Name & Description \\ \hline
    time        & Log timestamp in JST. \\
    uid         & User ID of the participant. \\
    classID     & Virtual room ID created by the organizer. \\
    taskID      & Task/question ID created by the organizer. \\
    lang        & Programming language used. \\
    op          & User action (e.g., compile, submit). \\
    msg         & Compile message (e.g., status, output, errors). \\ \hline
  \end{tabular}
\label{tab:table_C2Room}
\end{table}

\subsection{Data Source}
\label{section:data_source}
In this section, we explain the data source.
In this study, we use web browsing and programming logs as a data source.
We use the web browser logger \textit{TrackThinkTS} and \textit{C2Room}~\footnote{\url{https://C2Room.jp/}}, an online programming IDE for collecting programming activity logs. 
We will explain web browsing and programming logs in detail.

Web browsing activities are recorded through a web browser extension called \textit{TrackThinkTS}. Table~\ref{tab:table_TrackThinkTS} shows overall logs collected by the application. One notable advantage of \textit{TrackThinkTS} over other web browser loggers is its user-friendly interface, allowing participants to delete irrelevant logs collected during the experiment. Users need to install this extension in their web browser to utilize it. In our study, we have selected Google Chrome~\footnote{\url{https://www.google.com/chrome/}} as the preferred web browser.

Programming activities are recorded through an online web IDE called \textit{C2Room}. Table~\ref{tab:table_C2Room} shows overall logs collected by the application. The application is tailored for online programming classes in educational institutions and companies. The log lets us understand how and when users execute their code in the compiler and the compiled result. The application also allows users to track when they are satisfied with their written code and proceed to submit it.

\begin{table}[t!]
  \centering
  \caption{Selected log after from \textit{TrackThinkTS} and \textit{C2Room}.}
  \renewcommand{\arraystretch}{1.0}
  \small
  \begin{tabular}{l p{6.5cm}}\hline
    Column Name   & Description \\ \hline
    timestamp     & Time of the action in UNIXTIME. \\
    userID        & Participant ID during SRL. \\
    taskID        & Task/question ID created by the organizer. \\
    userAction    & Action logs collected from data sources. \\
    tabURL        & URL of the accessed web page. \\
    clipboardCopy & Text copied to the clipboard. \\
    msg           & Compile message (e.g., status, output, errors). \\ \hline
  \end{tabular}
\label{tab:table_data_fusion}
\end{table}

\subsection{Data Fusion}
\label{section:data_fusion}
Data fusion is applied to the collected data sources. The data fusion process comprises two primary procedures: data shaping and filtering. Data shaping involves transforming raw data into a unified format suitable for combining various data sources. On the other hand, data filtering entails selecting relevant data points following the conversion and concatenation. We will now provide a detailed explanation of each procedure.

\textbf{Data Shaping:} We employ data shaping as a preprocessing step to facilitate the concatenation and filtering of logs. We rename the columns for each web browsing and programming log. Specifically, we begin by renaming the columns \textit{date} and \textit{time} to \textit{timestamp}. Additionally, we rename \textit{UserID} and \textit{uid} to \textit{userID}. For \textit{UserAction} and \textit{op}, we change to \textit{userAction}. This column renaming process aims to align the corresponding information between the two applications. Once the column renaming is complete, we convert the \textit{timestamp} values to UNIXTIME and sort them chronologically.

\textbf{Data Filtering:} Table~\ref{tab:table_data_fusion} shows the table after data filtering. Some information is removed, such as window scroll speed from the web browsing log, class ID from the programming log, or logs involving NAN values. In order to sort logs into time order, we convert all units of the timestamp into UNIXTIME. Specifically, the programming logs collected by \textit{C2Room} were converted from JST to UNIXTIME. All the logs are concatenated and sorted in a time order using \textit{timestamp}.

\begin{figure}[t]
  \centering
  \includegraphics[width=0.8\linewidth]{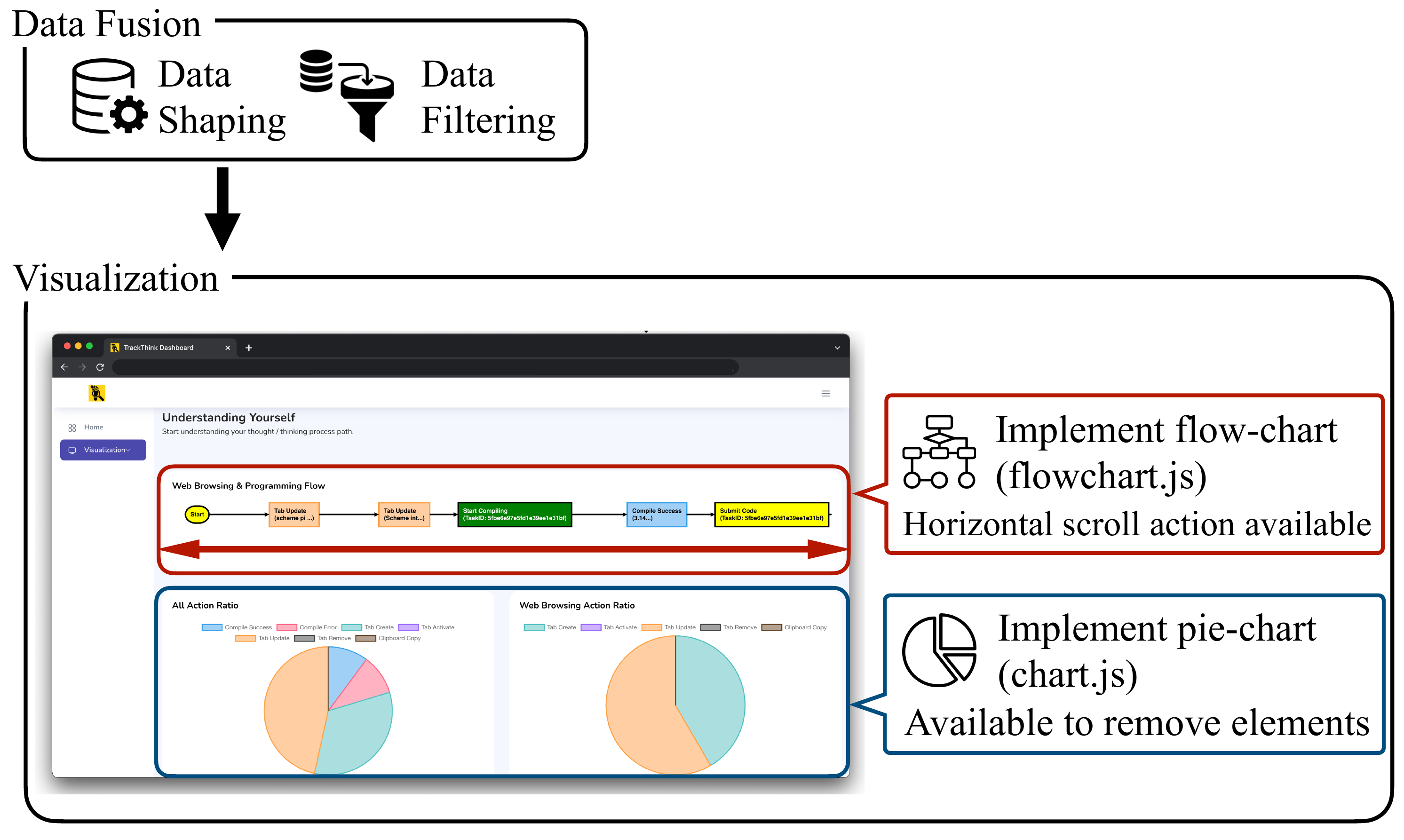}
  \caption{Application workflow. Visualization of flowchart and pie-chart.}
  \label{fig:visualization}
\end{figure}

\subsection{Visualize Web Browsing and Programming}
Visualization of web browsing and programming activity is performed after data fusion. Figure~\ref{fig:visualization} shows the library used for visualization. For the visualization, we choose a flowchart and pie-chart. We will explain the implementation process and the reason for the choice in detail.

The flowchart is selected to visualize the problem-solving progress in a time series.
The approach uses flowcharts to understand what search results participants used to arrive at their answers and what compilation errors they encountered when re-running their searches. The flowchart is implemented using \textit{flowchart.js}~\footnote{\url{http://adrai.github.io/flowchart.js}}.
It is a JavaScript library for flowchart SVG (Scalable Vector Graphics) rendering that runs in the terminal and browser. We categorize users' activities in different colors and shapes for start and stop edges. The workflow is not visualized fully on the screen, but the user can scroll horizontally to see the actions between the start and end edges. For edges like \textit{tab activate} or \textit{tab update}, the hyperlink is set so that users can jump to the webpage once they tap the edge.

The pie-chart is selected for visualization because the activity ratio of user action is essential in identifying users' domain knowledge~\cite{watanabe2022how}. The pie chart is implemented using \textit{chart.js}~\cite{da2019learn}. Clicking each element removes a specific activity from the pie chart. This option helps teachers or students focus on the ratio of an activity they want to compare. It is a JavaScript library for making HTML-based charts.

\section{Data Collection}
\label{section:data_collection}

\subsection{Participants}
In this experiment, we collect logs from lecture students (Dataset A) and non-lecture students (Dataset B).
The total number of participants is 33 unique (32 males and one female) university students in Japan.

Dataset A -- University students attending lectures: We collect data from 13 unique university students (12 males and one female) in Japan. 
Participants have taken university courses in Scheme programming language, so they have knowledge on Scheme grammar or syntax.

Dataset B -- University students not attending lectures: We collected data from 20 male university students in Japan. Participants did not take any university courses related to the Scheme, so they do not have any knowledge from the class, such as Scheme grammar or syntax.

\begin{figure}[t]
  \centering
  \includegraphics[width=0.4\linewidth]{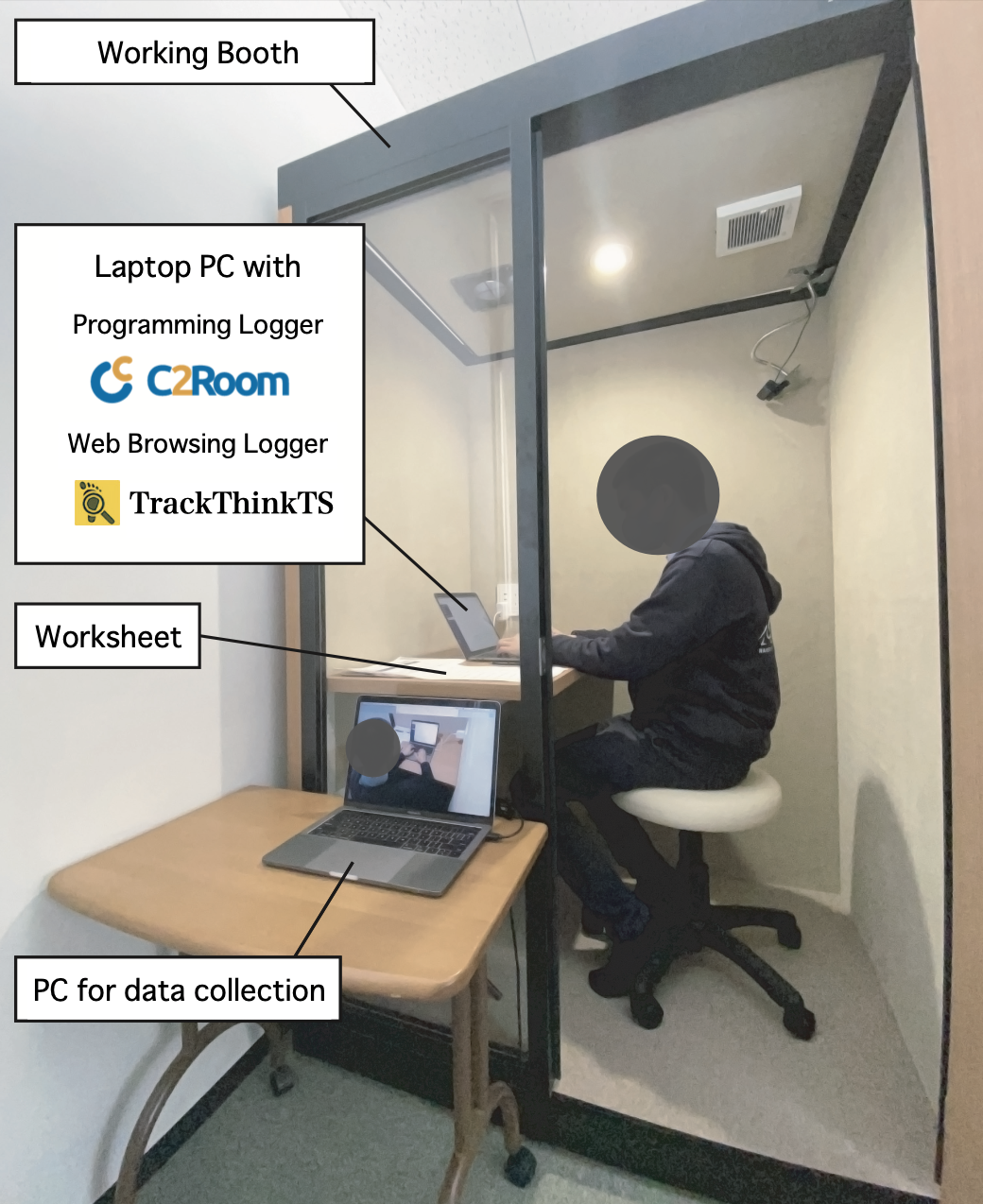}
  \caption{Experimental settings. Collect logs while solving scheme questions.}
  \label{fig:experimental-settings}
\end{figure}

\begin{table}[t!]
  \caption{Scheme Questions}
  \renewcommand{\arraystretch}{1.5}
  \resizebox{\columnwidth}{!}{\begin{tabular}{l l}
    \hline
      ID & Question \\ \hline
      A  & Define variable {\it PI} as 3.14.  \\
      B  & Write a scheme to show PI ${* 5^2}$. \\
      C  & Write a scheme to show ${(-b+\sqrt{b^2-3ac})/3a}$. \\
      D  & Define function {\it areaDisk} to calculate circle area from radius {\it r}. \\
      E  & Define function {\it areaRing} to calculate circle area from outer and inner diameter {\it D, d}. \\
      F  & Define function {\it d2y} that convert US currency dollar {\it d} to Japanese currency yen. \\
         & Note that 1 US dollar is 108.43 yen. \\
      G  & Define function {\it e2d} that convert. European currency {\it e} to United states dollar.\\
         & Note that 1 euro is 1.1069 US dollar. \\
      H  & Define function {\it p2e} that convert British currency pond {\it p} to European currency.\\
         & Note that 1 pond is 1.1632 euro. \\
      I  & Define function {\it p2y} that convert British currency pond {\it p} to Japanese currency.\\
         & Use {\it d2y}, {\it e2d}, and {\it p2e} in the previous questions. \\
      J  & Define function {\it c2f} that convert Celsius {\it C} to Fahrenheit. Note that ${f = 1.8c + 32}$. \\ \hline
  \end{tabular}}
  \label{tab:questions}
\end{table}

\subsection{\textbf{Experiment Procedure}}
Figure~\ref{fig:experimental-settings} shows the condition of the experiment. Before the experiment, \textit{C2Room} and \textit{TrackThinkTS} were installed on each participant's laptop. Under these conditions, the experiment was conducted using the following procedure: 
(1) The experiment conductor presents the purpose of the experiment, experimental conditions, and tools that will be provided to the participant. Only the person who agrees with the conditions can participate in the experiment. 
(2) The participant enters the personal workstation and starts web browsing and programming loggers. 
(3) Participants work on solving problems with a given schema using the \textit{C2Room} programming editor. 
Table~\ref{tab:questions} shows the questions for participant to solve.
The order of question-solving is not restricted, but it is assumed that the participant will solve the easy questions step by step. 
\textit{C2Room} will record the compiled results for each question. 
(4) Participants may use a web browsing to find the answer. \textit{TrackThinkTS} will track web browsing behavior. 
(5) The participant submits the answer and moves on to the next question when satisfied with the code compilation. Allow the student to return to previous questions to change the answer. 
(6) When all questions have been answered or an hour has passed, we stop participants from problem-solving. 
(7) Participants remove privacy-sensitive logs from \textit{TrackThinkTS}. Privacy-sensitive data refers to data stored during the experiment that is not related to the programming task or that the user does not want to share. Once all recorded logs are submitted to the experimenter, the participant leaves the personal workspace.

In this study, we recruit the same Scheme questions as \citet{watanabe2022how}. We have prepared ten questions in order of difficulty.
Easy questions are those that require fewer lines of code to solve. 
In this experiment, we chose Scheme (Racket), one of the dialects of LISP languages, as the programming language of the task~\cite{adams1995}. 
We chose this task because of its simple language specification, used in programming courses at several universities. 
Also, it is tailored for lecture use, so its language specification is usually unknown to students except those attending the lecture.

\section{Result}
In this section, we show results obtained from the \textit{TrackThinkDashboard}.

\begin{figure}[t!]
  \centering
  \includegraphics[width=1.0\columnwidth]{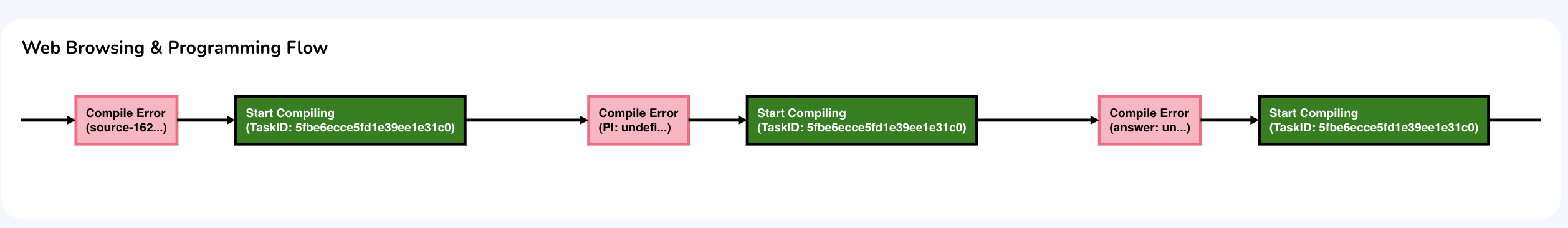}
  \caption{\textbf{Try-and-error student}: Student receive an error response after compiling, and students try to compile before going back to the web search.}
  \label{fig:try_and_error_group}
\end{figure}

\begin{figure}[t!]
  \centering
  \includegraphics[width=1.0\columnwidth]{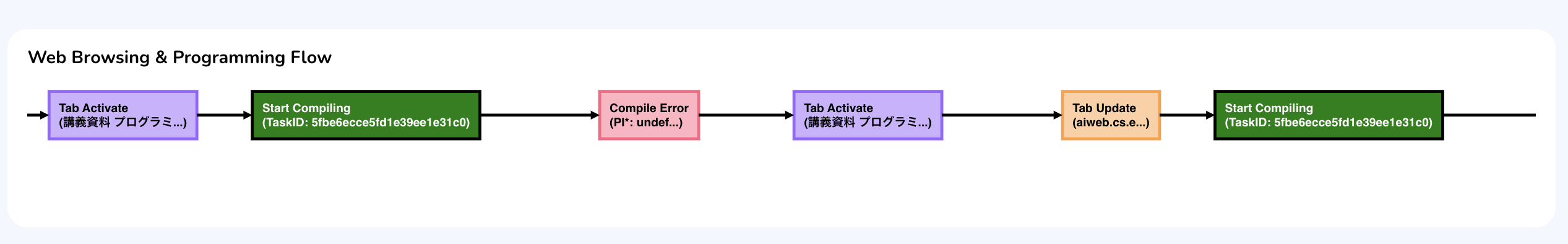}
  \caption{\textbf{Try-and-search student:} Student receive an error response after compiling, and students return to the web browsing to find a solution before the following compilation.}
  \label{fig:try_and_search_group}
\end{figure}

\begin{figure}[t!]
  \centering
  \includegraphics[width=1.0\columnwidth]{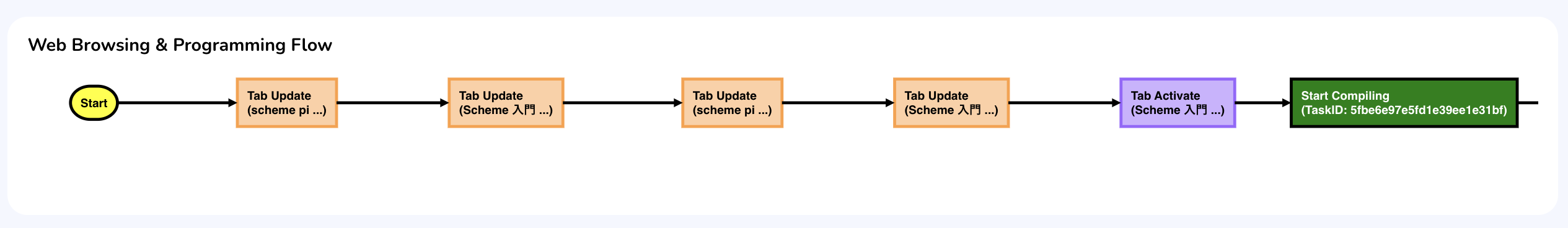}
  \caption{\textbf{Cautious student}: Student use web searches before programming. Once the solution is identified, write code from scratch or use a clipboard copy to solve the task.}
  \label{fig:cautios_group}
\end{figure}

\begin{figure}[t!]
  \centering
  \includegraphics[width=1.0\columnwidth]{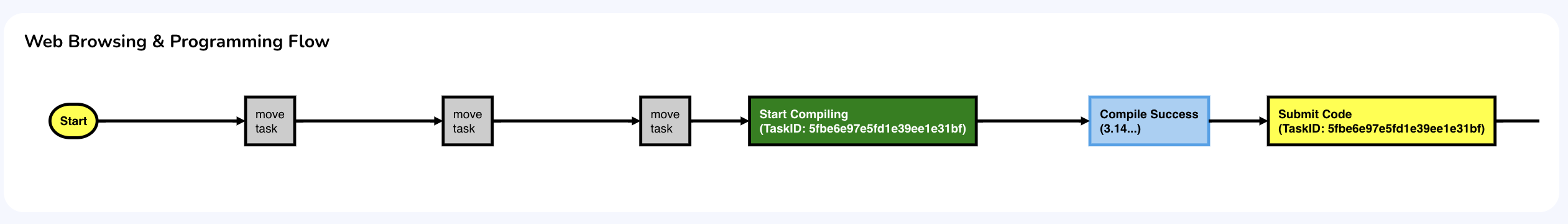}
  \caption{\textbf{Time management student}: Student move to other tasks after starting problem-solving. One participant look for a few questions, whereas the other student look at all questions before starting to solve questions.}
  \label{fig:time_manager_group}
\end{figure}

Figure~\ref{fig:try_and_error_group} shows a sample student workflow for solving a programming problem using the try-and-error approach. Student compile the code, and after receiving an error response, they modify it and try to recompile it.

Figure~\ref{fig:try_and_search_group} shows an example student working on the same task and receiving an error response, but this user decides to go back to the web search before the following compilation. We call this pattern of solving a try-and-search student. The student copies the error message from the compiler and inserts and searches in the web browser. Both students receive an error response on the same problem, but each student acts differently to continue solving the problem.

Figure~\ref{fig:cautios_group} illustrates a sample workflow of a student group tackling a programming task. A defining characteristic of these students is their approach of first gathering information from web pages before attempting to solve the problem. Once they have a clear understanding of the solution, they return to the online IDE to code and submit their work. While two students in this group follow similar workflows, one writes the code from scratch, whereas the other copies sections of code using the clipboard. Both demonstrate a methodical approach, prioritizing understanding over immediate action. We refer to this group as \textit{cautious students}.

\begin{figure}[t!]
  \centering
  \includegraphics[width=1.0\columnwidth]{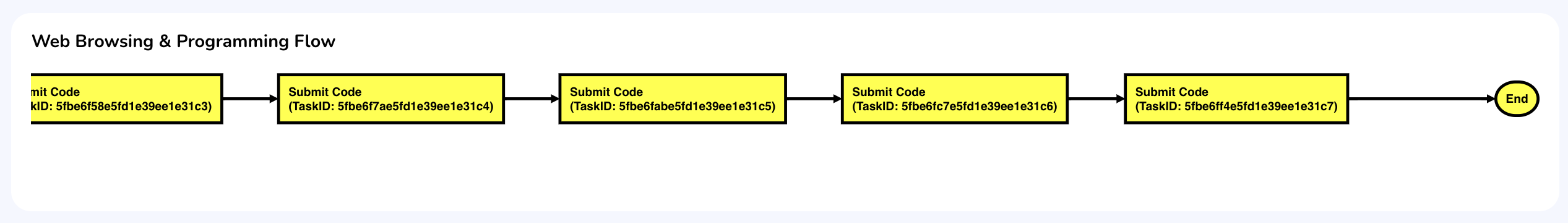}
  \caption{\textbf{Double checking student:} Student move to previous questions to double check and submit their code before finishing the experiment.}
  \label{fig:double_check_group}
\end{figure}

\begin{figure}[t!]
  \centering
  \captionsetup[subfigure]{justification=centering}
  \subfloat[Sample pie-chart for non-lecture attendee student]{
    \includegraphics[width=0.33\columnwidth]{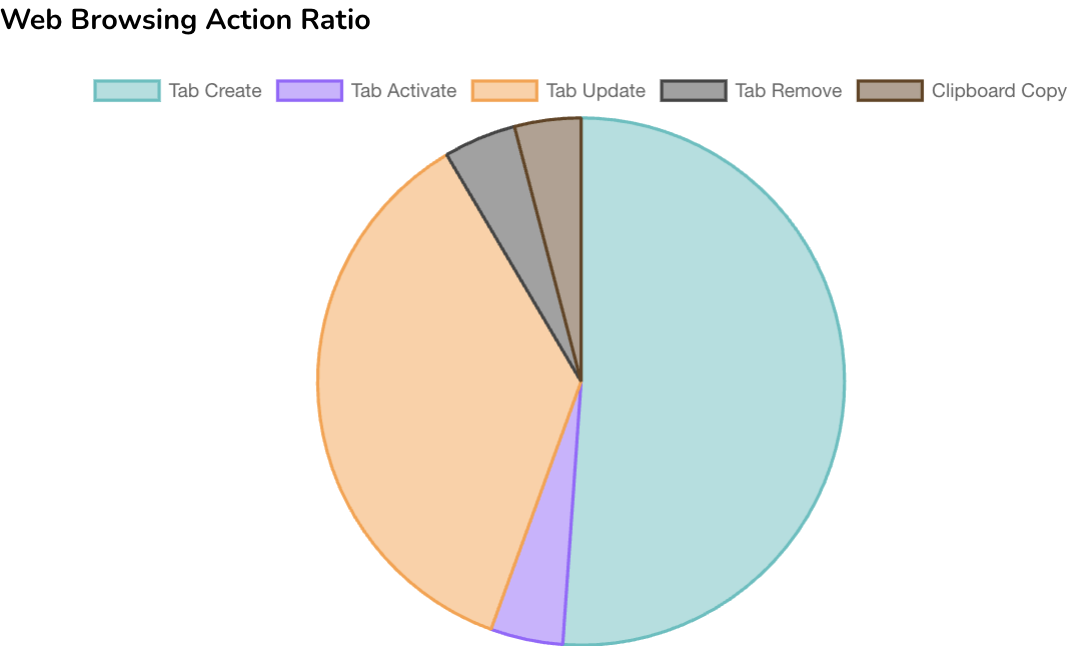}
    \includegraphics[width=0.27\columnwidth]{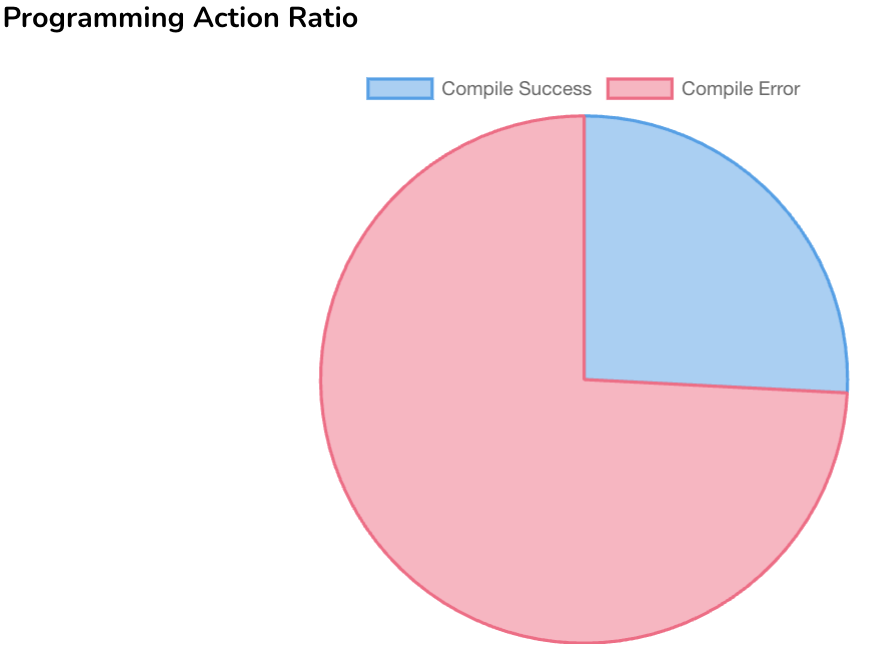}
    \includegraphics[width=0.33\columnwidth]{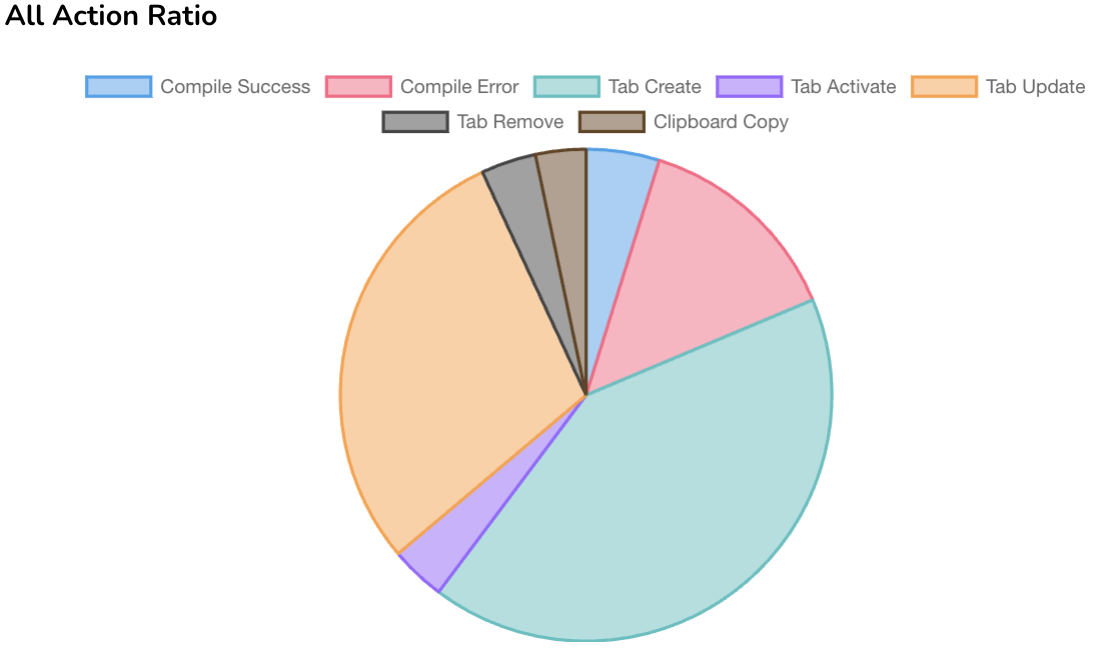}
    \vspace{0.2cm}
  }
  \vspace{0.6cm}
  \subfloat[Sample pie-chart for lecture attendee student]{
    \includegraphics[width=0.33\columnwidth]{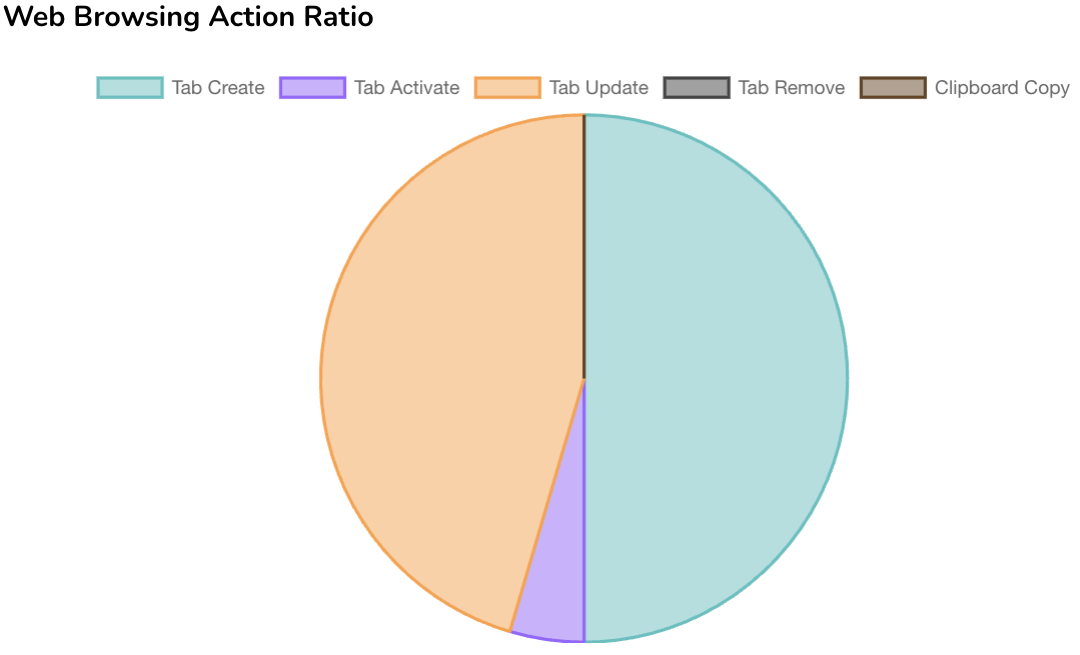}  
    \includegraphics[width=0.27\columnwidth]{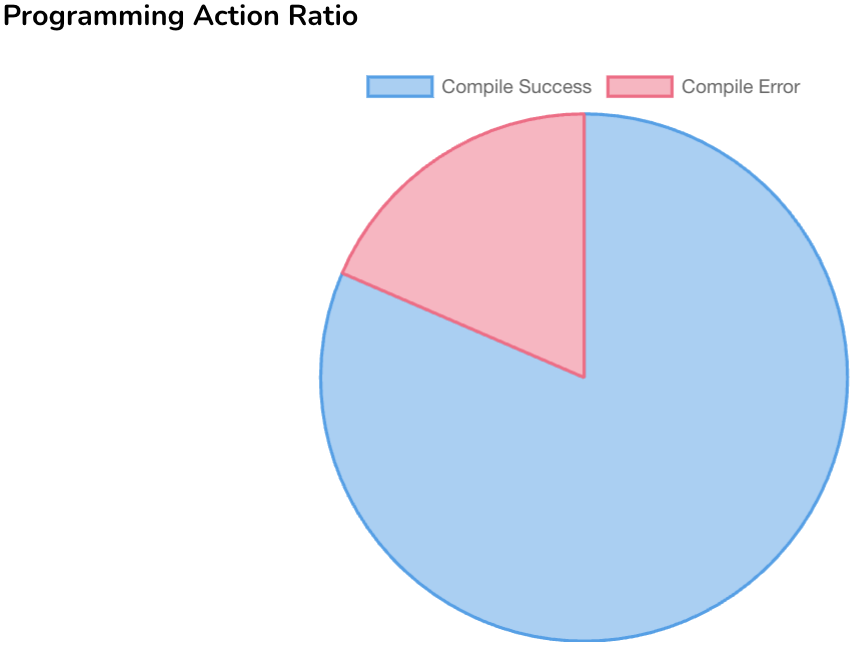}
    \includegraphics[width=0.33\columnwidth]{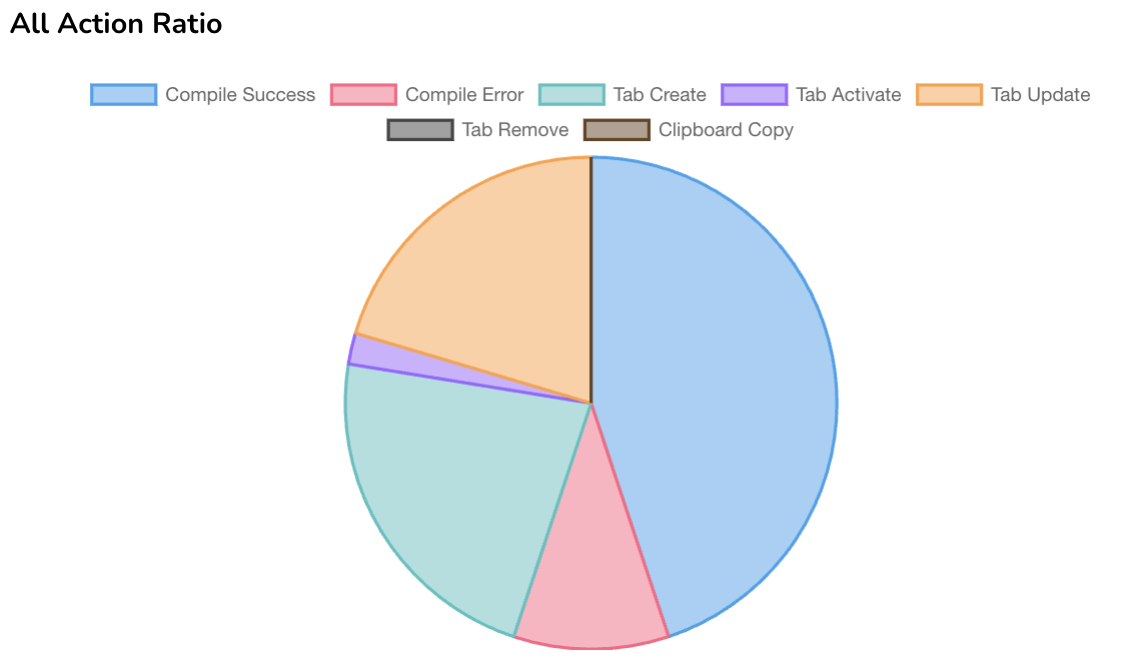}
    \vspace{0.2cm}
  }
  \caption{Pie-chart of selected non-lecture and lecture attendance students. The pie-chart represents a ratio of students' action counts. The left shows an action ratio of web browsing logs, the middle shows the ratio of programming compilation result counts, and the right shows the ratio of all action counts.}
  \label{fig:pie_chart}
\end{figure}

Figure~\ref{fig:time_manager_group} presents an example workflow of students who strategically review multiple questions at the start of the experiment. While one student examines only a few questions, another reviews all of them before beginning their solutions. Notably, this behavior was absent among non-lecture students and was exclusive to lecture attendees, who possess domain knowledge of the programming language. These students prioritize solving easier questions first, optimizing their time and effort. We refer to this group as \textit{time management students}.

Figure~\ref{fig:double_check_group} shows a workflow typical of the \textit{double-checking student}. Before concluding the experiment, student revisit all programming questions to carefully review their submissions. This behavior reflects their thorough and detail-oriented problem-solving approach.

Finally, Figure~\ref{fig:pie_chart} displays action ratio pie charts from the experiment, comparing non-lecture and lecture attendees. The pie charts illustrate the proportion of time spent on web browsing, programming, and a combination of the two. Students who attended lectures, possessing greater domain knowledge, exhibited a higher success rate in code compilation compared to errors. Conversely, non-lecture students experienced more compilation errors and relied more heavily on web searches. These differences align with findings from prior research~\cite{watanabe2022how}.

\section{Discussion}
\label{section:discussion_future_work}
Our study visualizes a flowchart by combining web browsing and programming data to understand students' self-regulated learning (SRL) workflows. This approach provides insights into how students tackle each task, analyzing their learning activities while uncovering individual differences. Flowcharts capture detailed problem-solving processes, while pie charts highlight frequently used actions, offering a clearer understanding of programming language comprehension~\cite{watanabe2022how}. Pie charts help monitor problem-solving progress, while flowcharts enable a closer examination of the steps students take, shedding light on their thought processes and strategies.

Although effective, the current visualization has limitations. One limitation is the lack of action duration representation. Actions are sorted by timestamps, but varying node lengths to reflect duration could help identify when students encounter difficulties. The flowchart could also be enhanced by separating actions into branches, such as web browsing and programming, inspired by Git's branching system~\footnote{\url{https://git-scm.com/docs/git-branch}}.
While the current single-line design is simpler, branching could reveal more nuanced workflows in some contexts.

Integrating additional sensor data offers another improvement opportunity. Potential plugins include wearable devices to monitor physiological data for measure stress~\cite{garg2021stress} or fatigue~\cite{ElsenRonando202418}, eye-tracking~\cite{AnkurBhatt20249} to measure attention~\cite{ishimaru2016towards} or concentration~\cite{SakiTanaka202410}, facial recognition technology to assess micro-behaviors or engagement levels~\cite{watanabe2021discaas, watanabe2023engauge}, and bluetooth based position estimation~\cite{bello2024besound}. Such data could provide deeper insights into students’ cognitive states and behaviors.

Our analysis identified five SRL patterns during web browsing and programming: \textit{try and error}, \textit{try and search}, \textit{cautious}, \textit{time management}, and \textit{double checking}. Combining knowledge input (web browsing) with knowledge output (programming) revealed unique problem-solving approaches that submitted code alone cannot capture. These insights help teachers understand students better and guide their coaching. Additionally, the dashboard highlights high-performing students' workflows, offering a resource for novice learners. Pie charts also reveal domain knowledge~\cite{watanabe2022how}.

\section{Limitation and Future Work}
In this study, several limitations remain.
Our study works on visualizing programmers' web browsing and coding behavior in the sequential workflow. We discovered the difference in the student's approach to programming. However, this study lacks direct comparisons between students who used the system and those who did not. In our future study, we will evaluate students' self-regulated learning performance.

Another limitation is the lack of integration of real-time feedback mechanisms and the collection of qualitative data on user experiences. Our work may allow us to visualize the programming and web browsing activities, but we did not implement the system work in real-time; instead, we did post-visualization. Our future work should consider collecting students' SRL abilities over time, and better to receive results in real-time, which then the lecturer can interact with a student for supervising how to improve in programming learning.

The limitation also aligned with the data collection. In this study, we recruited 33 unique (32 males and one female) university students in Japan. The characteristics we observe in the student groups are still a range of Japanese and mainly male. Our future work will be to recruit more participants to discover different types of web browsing and programming behavior.

Other then that, current analysis focuses on individual patterns without considering combinations, such as \textit{cautious} students also using \textit{time management} strategies. Exploring such combinations could provide a better understanding of problem-solving processes. The programming tasks used in this study were relatively simple, focusing on Scheme syntax, leading to limited score variability. Future work should design tasks that encourage diverse strategies and performance levels.

Lastly, overcoming the privacy constraints is another future work. To avoid risk of personal browsing data, we have not made the raw dataset publicly available. However, we are exploring ways to anonymize the data so that the web browsing and coding logs could be shared for reproducibility. We anticipate future releases of partial data under an institutional review framework.

\section{Conclusion}
\label{section:conclusion}
In this paper, we introduced \textit{TrackThinkDashboard}, a system that visualizes self-regulated learning (SRL) workflows via flowcharts and pie charts to capture both web browsing and programming activities. By collecting data from 33 university students (with varying backgrounds in Scheme), our dashboard revealed diverse problem-solving patterns that spotlight how students allocate effort between information-seeking and code revisions. Beyond mere identification of learning strategies, these visualizations explicitly support teachers by uncovering when and why certain learners become stuck, and by highlighting frequent error–search cycles or effective workflows from high-performing students. Armed with this data, teachers can provide more targeted guidance—for instance, by reviewing the specific resources a struggling student consults or offering additional examples for recurring error messages. Novice learners can also study the flowcharts of advanced peers to adopt more efficient approaches, such as balancing trial-and-error with timely research on relevant syntax or debugging techniques. Taken together, \textit{TrackThinkDashboard} offers a practical framework for both instructors—who can rapidly identify at-risk students and tailor feedback—and students aiming to refine their meta-cognitive awareness. Future work will focus on integrating these visualizations into ongoing classroom use, refining the data pipeline for larger-scale or real-time deployments, and evaluating long-term impacts on SRL skill development. We have also addressed manuscript repetition, clarified the paper’s scope and audience, and corrected template formatting to better align with publication requirements.

\section*{Acknowledgement}
This work is partially supported by JSPS KAKENHI Grant Number 23KK0188.

\bibliographystyle{plainnat}
\bibliography{Authors/bibtex}
\end{document}